\documentclass[aps,prb,superscriptaddress,reprint,floatfix]{revtex4-1}
\usepackage{amsmath}
\usepackage{graphicx}
\usepackage{longtable}
\usepackage{colortbl}
\usepackage[english]{babel}
\usepackage{array}
\usepackage{booktabs}

\graphicspath{{graphs/}}

\begin{document}

\title{Probe Ferroelectricity by X-ray Absorption Spectroscopy in Molecular Crystal}
\author{Fujie Tang} \affiliation{Department of Physics, Temple University, Philadelphia, PA 19122, USA}
\author{Xuanyuan Jiang} \affiliation{Department of Physics and Astronomy, Nebraska Center for Materials and Nanoscience, University of Nebraska, Lincoln, Nebraska 68588, USA}
\author{Hsin-Yu Ko} \affiliation{Department of Chemistry, Princeton University, Princeton, NJ 08544, USA}
\author{Jianhang Xu} \affiliation{Department of Physics, Temple University, Philadelphia, PA 19122, USA}
\author{Mehmet Topsakal}\thanks{$^\ast$Current address: Nuclear Science and Technology Department, Brookhaven National Laboratory, Upton, New York 11973, USA}
\affiliation{Center for Functional Nanomaterials, Brookhaven National Laboratory, Upton, New York 11973, USA}

\author{Guanhua Hao} \affiliation{Department of Physics and Astronomy, Nebraska Center for Materials and Nanoscience, University of Nebraska, Lincoln, Nebraska 68588, USA}
\author{Alpha T. N’Diaye} \affiliation{ Advanced Light Source, Lawrence Berkeley National Laboratory, Berkeley, California 94720, USA}
\author{Peter A. Dowben} \affiliation{Department of Physics and Astronomy, Nebraska Center for Materials and Nanoscience, University of Nebraska, Lincoln, Nebraska 68588, USA}
\author{Deyu Lu} \affiliation{Center for Functional Nanomaterials, Brookhaven National Laboratory, Upton, New York 11973, USA}
\author{Xiaoshan Xu}
\affiliation{Department of Physics and Astronomy, Nebraska Center for Materials and Nanoscience, University of Nebraska, Lincoln, Nebraska 68588, USA}
\author{Xifan Wu}
\affiliation{Department of Physics, Temple University, Philadelphia, PA 19122, USA}
\date{\today}

\begin{abstract}
\par We carry out X-ray absorption spectroscopy experiment at oxygen K-edge in croconic acid (C$_5$H$_2$O$_5$)
crystal as a prototype of ferroelectric organic molecular solid, whose electric polarization is generated by
proton transfer. The experimental spectrum is well reproduced by the electron-hole excitation theory
simulations from configuration generated by {\it ab initio} molecular dynamics simulation. When inversion
symmetry is broken in ferroelectric state, the hydrogen bonding environment on the two bonded
molecules become inequivalent. Such a difference is sensitively probed by the bound excitation
in the pre-edge, which are strongly localized on the excited molecules. Our analysis shows that a
satellite peak in the pre-edge will emerge at higher excitation energy which serves as a clear
signature of ferroelectricity in the material.
 \end{abstract}

\maketitle

\section{INTRODUCTION}

\par Ferroelectricity (FE) is an important material property describing the ability of some solids to sustain a spontaneous electric polarization, whose direction can be reversed under applied electric fields \cite{Lines2001,Dawber2005,Rabe2007,Scott2007,Xu2015,Liu2016}. Because of its importance in both fundamental science and technology, extensive techniques have been developed to study the nature of FE since its discovery one century ago. In experiments, the polarization profile is sensitive to the optical second harmonic generation \cite{Denev2011}, and the electric dipole density in crystals can be quantitatively determined in the hysteresis measurement through the flow of the polarization current \cite{Lee2005}. Spectroscopy experiment, such as X-ray absorption spectroscopy (XAS), is also widely applied in solids \cite{Kholkin2007,Iwano2017c}. In XAS process, the timescale of electron-hole excitation is much shorter than that of the lattice vibrations. Therefore, XAS spectra record the electronic structural information corresponding to its instantaneous local environment \cite{Wernet2004a,Chen2010,Prendergast2006b,Bisti2011d,Sun2017,Sun2018a}. However, in conventional oxide ferroelectrics, XAS spectra were only employed to indirectly infer the underlying FE orderings since the degree of hybridization between transition metal and oxygen is only slightly affected by the polar distortion \cite{Ravel1995,Cabrera2011}. Unfortunately, an unambiguous XAS signature of FE so far remains elusive, which requires more abrupt electronic structure changes induced by broken inversion symmetry. 

\par Recently, the successful synthesis of hydrogen (H)-bonded organic ferroelectrics has attracted much attention \cite{Horiuchi2008,Horiuchi2010,Horiuchi2012,Horiuchi2017,Horiuchi2018,Owczarek2016,Stroppa2011c,Jiang2016}. In FE molecular crystals, such as croconic acid (CA), the measured polarizations are comparable to that of BaTiO$_3$ as one of the most studied ferroelectrics \cite{King-Smith1994,Zhong1994}. Molecular ferroelectrics exhibit at much lower switching fields than oxide ferroelectrics \cite{Horiuchi2017,Horiuchi2008,Horiuchi2010,Horiuchi2012}. Such intriguing physics originates from the unusual way in generating FE. For CA, the polar distortion arises from the large displacements of protons, which is conceptually described as the {\it proton transfer} mechanism \cite{Horiuchi2017,Cai2013,Iwano2017c}. When a proton is located at the middle point of two oxygen atoms along an H-bond, the two molecules formed a symmetric H-bond environment.  The polarization emerges as long as protons displace away from the high symmetry point along the H-bonding direction \cite{DiSante2012b}. Besides, the protons also make significant electronic contribution to the overall polarization due to the large charge transfer accompanied by proton transfer. Such a large electronic structure changes via proton transfer in molecular crystals also provides an ideal platform where FE can be probed by XAS. In the organic ferroelectrics, the two bonded molecules no longer have the same H-bonding strength once proton transfer occurs. The resulting two distinct H-bonding configurations break the inversion symmetry and should have clear signatures in XAS spectrum. Indeed, XAS at the oxygen K-edge has recently been applied as a sensitive probe for the H-bonding environment in water \cite{Wernet2004a, Tse2008, Chen2010, Prendergast2006b, Sun2017,Sun2018a}.

\par Here, we address the above issues by combined experimental and theoretical studies of oxygen $K$-edge XAS in CA molecular crystal. Based on the electron-hole excitation theory \cite{Prendergast2006b}, the computed XAS spectrum agrees well with our experimental one at room temperature. Our electronic structural analyses further reveal that proton transfer not only generates
a large electric polarization that strongly couples to the $\pi$ bonding state, but also lifts the degeneracy of the excitation with $\pi^{\ast}$ antibonding character represented by the pre-edge of XAS. With the development of FE via proton transfer, two equivalent H-bonded molecules evolve into H-bond donor and H-bond acceptor molecules, respectively. Influenced by the Coulomb potential of the proton, the $\pi^{\ast}$ exciton on the H-bond donor molecule becomes delocalized with an increased excitation energy; and the opposite effect is observed for the $\pi^{\ast}$ exciton on the H-bond acceptor molecule. Therefore, the resulting split pre-edge features at $\sim 530$ eV and $\sim 532$ eV respectively carry out unambiguous spectral signature of the broken inversion symmetry. Our method provides a new probe for FE via proton transfer mechanism, which is ready to be applied on other molecular ferroelectrics in general.

\section{METHODS}
\label{part2}
\par The polycrystalline CA films (450 nm) were deposited on $10\times10$ $\rm {mm^2}$ highly ordered pyrolytic graphite substrates using physical vapor deposition in high vacuum ($\rm {1.0 \times 10^{-7}}$ Torr), with a growth rate of 0.1 \AA/s. The film morphology and thickness were characterized by atomic force microscopy and the root mean square roughness was found to be about 16 nm \cite{Jiang2016}. The XAS spectra were carried out at the bending magnet beamline 6.3.1, in the Advanced Light Source of the Lawrence Berkeley National Laboratory. The photon flux was estimated to be $\rm {1.65 \time 105}$ $\rm {photons/(s\cdot\rm{\mu m^2})}$. The XAS spectra were taken at 300 K, in the total electron yield mode, within the energy range around the O K edge. Note that the Curie temperature of CA is higher than the decomposition temperature\cite{Horiuchi2010}. 

\par The theoretical predictions of the structure of the CA crystal were performed at both $T$ = 0, 300 K based on density functional theory (DFT) \cite{Hohenberg1964,Kohen1965} and {\it ab initio} molecular dynamics \cite{Car1985,marx2009} under isothermal-isobaric ensemble (constant $NpT$) \cite{Parrinello1980,BERNASCONI1995501} as implemented in Quantum Espresso \cite{Giannozzi2017a}. The LDA, the GGA in the form of PBE, SCAN (meta-GGA), and SCAN0 (hybrid-meta GGA, which mixed with 10$\%$ Hartree-Fock exchange)  \cite{Staroverov2003,Sun2015} to describe the electron exchange and correlation using a linear-scaling algorithm \cite{Wu2009,DiStasio2014} based on maximally localized Wannier functions \cite{Marzari2012,Marzari1997}. We used energy cut-off of 200 Ry to get the better convergence of lattice parameters with Monkhorst-Pack $4 \times 6\times 3$ k-points sampling. All atom positions were fully relaxed until the force on each atom was less than $10^{-6}$ a.u.. Furthermore, we carried out the Born-Oppenheimer Molecular Dynamics (BOMD) simulations with NpT ensemble for LDA, PBE, and SCAN functional with initial relaxed structures, while the Car-Parrinello molecular dynamics (CPMD) simulations with NpT (isothermal-isobaric) ensemble for SCAN0 functional with initial relaxed structures by using Quantum Espresso.  The NpT simulation is performed by using the Parrinello–Rahman barostat as implemented in Quantum Espresso code package.  The choice for using BOMD instead of CPMD with LDA, PBE and SCAN functionals is owing the unstable NpT simulation because of the relative underestimated band-gap of LDA and semi-local functional. The simulation cell contains 8 croconic acid molecules. The temperature is set at 300 K by using single Nose–Hoover chain thermostat. The time step for equation of motion is set to 2.0 a.u. for CPMD and 20.0 a.u. for BOMD. We adopted the norm-conserving Vanderbilt pseudopotential for C, O and H atoms and set the kinetic energy cutoff of electronic wavefunction as 130 Ry. We equilibrated the system with 2 ps, which has been proved to be enough for molecular crystal. Then we performed CPMD simulation production run up to 7 ps for LDA, PBE and SCAN functionals, while the length of trajectory for production run with SCAN0 is 5 ps. We picked up one snapshot from SCAN0 trajectory to perform XAS calculation.

The theoretical XAS spectrum was computed based on our recently developed computational method in treating electron-hole interactions \cite{Prendergast2006b,Chen2010,Sun2017,Sun2018a}. In order to sample the different local environment of oxygen atom, we excited all of 40 oxygen atoms in the cell where the snapshot is taken from trajectory of AIMD+SCAN0. Since we are focusing on the molecular crystal, even for the room temperature, one snapshot is more than enough to converge the XAS spectra. The theoretical structure and polarization predicted by AIMD simulation at the SCAN0 level are superior compared with the ones based on commonly used semi-local functionals.

\section{RESULTS and DISCUSSIONS}
\label{sis}

\begin{figure}[t!]
	\includegraphics[width=2.85in]{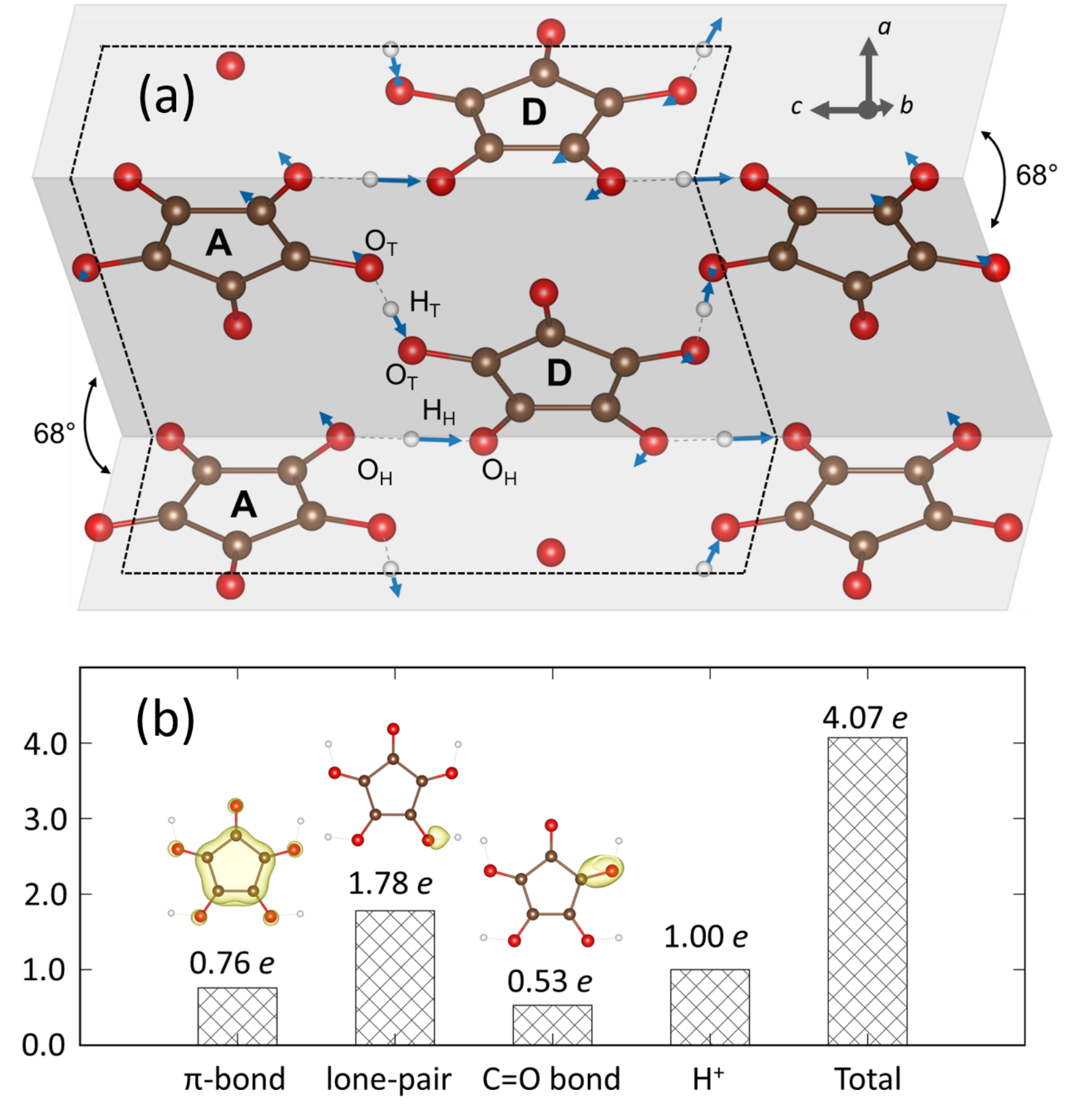}
	\caption{\label{fig:structure}
		(a) The schematic of the CA crystal unit cell and the vector symbols denote the structure distortion. (b) Decomposition of Born effective charge of the {\it hinge} H atom onto the contributions from Wannier functions of various bonding characteristics.}
\end{figure}

\par The CA molecule is composed of pentagon of carbon atoms, each of which is bonded to an oxygen atom. In condensed phase, its molecular alignment is governed by the intermolecular forces provided by the H-bond. As shown in Fig. 1(a), the two molecules within the same layer is connected by the {\it terrace} H-bond, and the two neighboring layers are linked by the {\it hinge} H-bond with a folding angle around $\rm{68^\circ}$. In crystals, the above folding layers results in zigzag shapes extending along three crystal directions. At room temperature, the corresponding lattice constants and volume are determined respectively to be $a = 8.62(8.50)\rm\AA$, $b = 5.10(5.18)\rm\AA$, $c = 10.92(10.82) \rm\AA$, and $V = 479.5(476.4)\rm \AA^{3}$. In the above, the numbers outside (in) the parentheses are our experimental (theoretical) values. At T=300 K, our SCAN0-AIMD simulation in croconic acid crystal yields an average electric polarization of 30.30 $\rm \mu C/cm^2$, which is close to the experimentally measured polarization of 30 $\rm \mu C/cm^2$ \cite{Horiuchi2012}. It should be stressed that the hybrid functional SCAN0 at the state-of-art meta-GGA level gives superior predictions of the H-bonded molecular structures than the conventional functional approximations.

The vector symbols in Fig. 1(a) denote the structural distortions in CA crystal from its hypothetical paraelectric phase to the FE ground state, with Pbcm and $\rm{Pca2_1}$ space group symmetry respectively \cite{Horiuchi2010,DiSante2012b,Cai2013,Kroumova2003}. Clearly the distortions are dominated by proton transfer as evidenced by the large displacements of protons along the H-bonding directions threading oxygen atoms on two neighboring molecules. The proton transfer is accompanied by a buckling mode of the oxygen atoms that slightly twists the pentagons \cite{DiSante2012b}. While the protons are in the middle of the two neighboring molecules, the crystal is invariant under a reflection operation, therefore the FE is forbidden. Once the proton moves away from the middle point, the inversion symmetry is broken, leaving one of the molecules as the H-bond donor and the other as the H-bond acceptor in Fig. 1(a). The proton transfer not only induces the polarity but also plays a key role in stabilizing the FE. For protons along either the {\it hinge} or the {\it terrace} direction, the computed dynamic effective charge tensor from linear response theory is found to have a very large diagonal element, the value for the {\it hinge} H atom is $\rm{Z_{H, PT}^*}=4.1 {\it e}$ along the proton transfer (PT) direction \cite{King-Smith1993,Zhong1994,Sharma2005}. Whereas the other diagonal and off-diagonal elements of the effective charge tensor are relatively small. The much larger effective charge than its nominal charge is not unusual in FE materials. It indicates that FE is promoted by the charge transfer under proton transfer by favoring the long-range Coulomb interaction over the short-range repulsion \cite{Ghosez1996,Samara1975a}. Based on modern theory of polarization \cite{Marzari2012,Marzari1997}, we further decompose $\rm{Z_{H, PT}^*}$ into the contributions from centers of maximally localized Wannier functions as shown in Fig. 1(b). As expected, $\rm{Z_{H, PT}^*}$ is mainly contributed by the Wannier functions of the lone pair electrons, which describes the local electronic structural changes in the process of proton transfer. Surprisingly, a significant electronic contribution in $\rm{Z_{H, PT}^*}$ is also identified to originate from the Wannier functions of the $\rm{\pi}$-bond on the H-bonded molecules in Fig. 1(b). It suggests that the ionic displacement of the proton transfer is actually strongly coupling to the $\rm{\pi}$-bond.  Indeed, it has long been proposed that the FE switching process largely involves the delocalized $\rm{\pi}$-bond in these organic FE materials \cite{Horiuchi2010,Horiuchi2017}.


\begin{figure}[t!]
	\includegraphics[width=3.1in]{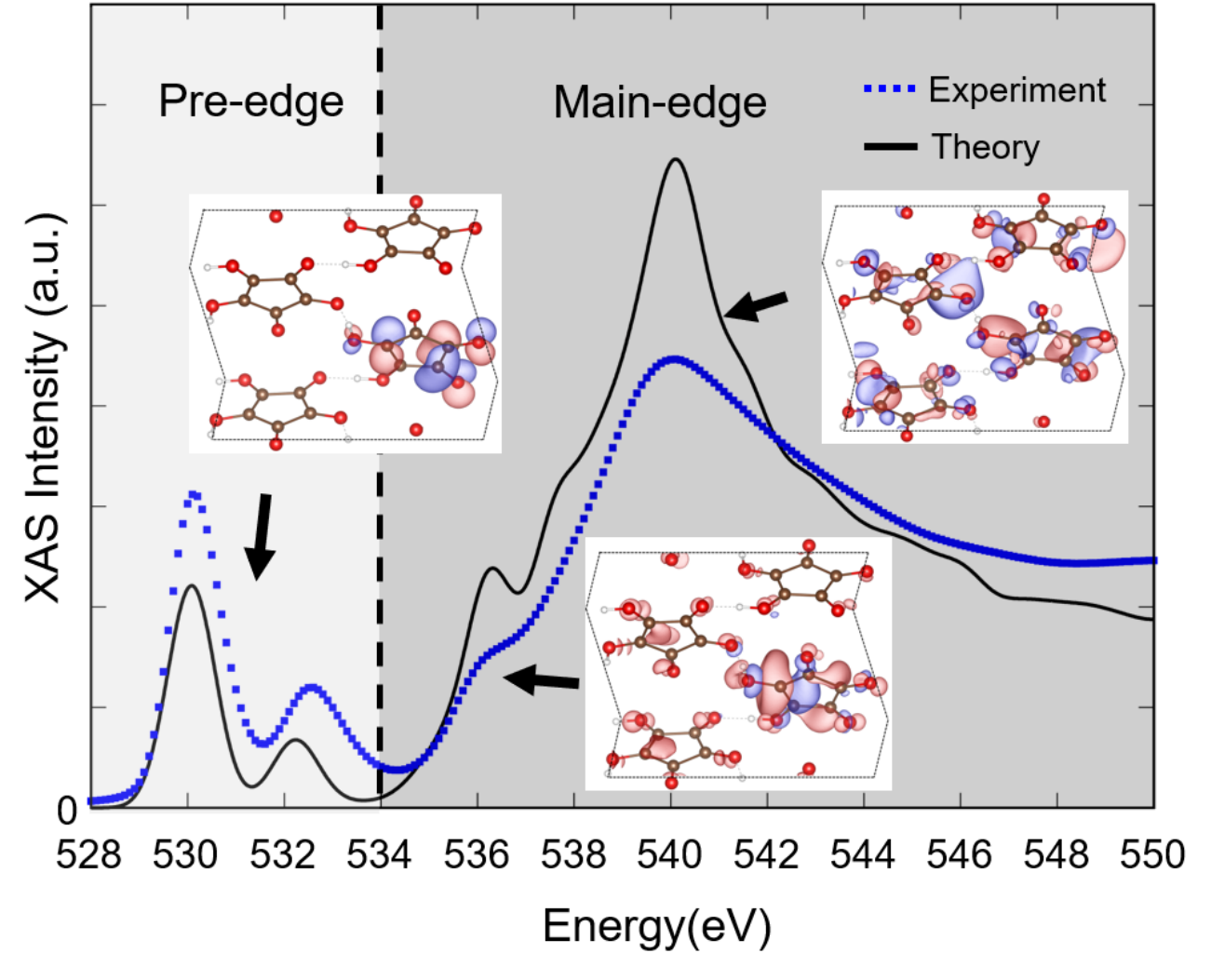}
	\caption{\label{fig:XAS_overall}
		The experimental (blue dashed line) and theoretical (black solid line) near edge X-ray absorption spectra of CA crystal. The theoretical XAS spectra was generated by using the configuration from {\it ab initio} molecular dynamics with SCAN0 at 300 K. The inserted figures are the representative orbital for pre-edge, kink and main-edge features of XAS. The excited oxygen atom is located in the downright CA molecule.}
\end{figure}

\par The broken inversion symmetry should not only be reflected in the electric properties of the
bonding electrons, but also leave a trace on the electron-hole excitations with
antibonding characteristic which could be probed by XAS. As shown in Fig.~\ref{fig:XAS_overall},
we present near the edge X-ray absorption fine structure of the CA crystal.
Within the $\sim$20 eV range, a good agreement can be seen
between experiment and theory in terms of both spectral intensity and line shape.
The pre-edge feature of XAS is located at relatively lower energies from 528 eV to 534 eV.
In the pre-edge, one sharp peak can be identified centered at 530.1 eV from experiment (530.1 eV from theory),
which is followed by a less pronounced feature at a slightly higher energy of 532.6 eV (532.1 eV).
Compared to the pre-edge feature, the main-edge of XAS from 534 eV to 548 eV is comprised of
broader spectral signals with stronger intensities centered around 540 eV (540 eV).
Moreover, a kink feature can also be seen at 536.1 eV (536.1 eV) on the left shoulder of the main-edge in XAS.

\par
\begin{figure*}[!htb]
	\includegraphics[width=7in]{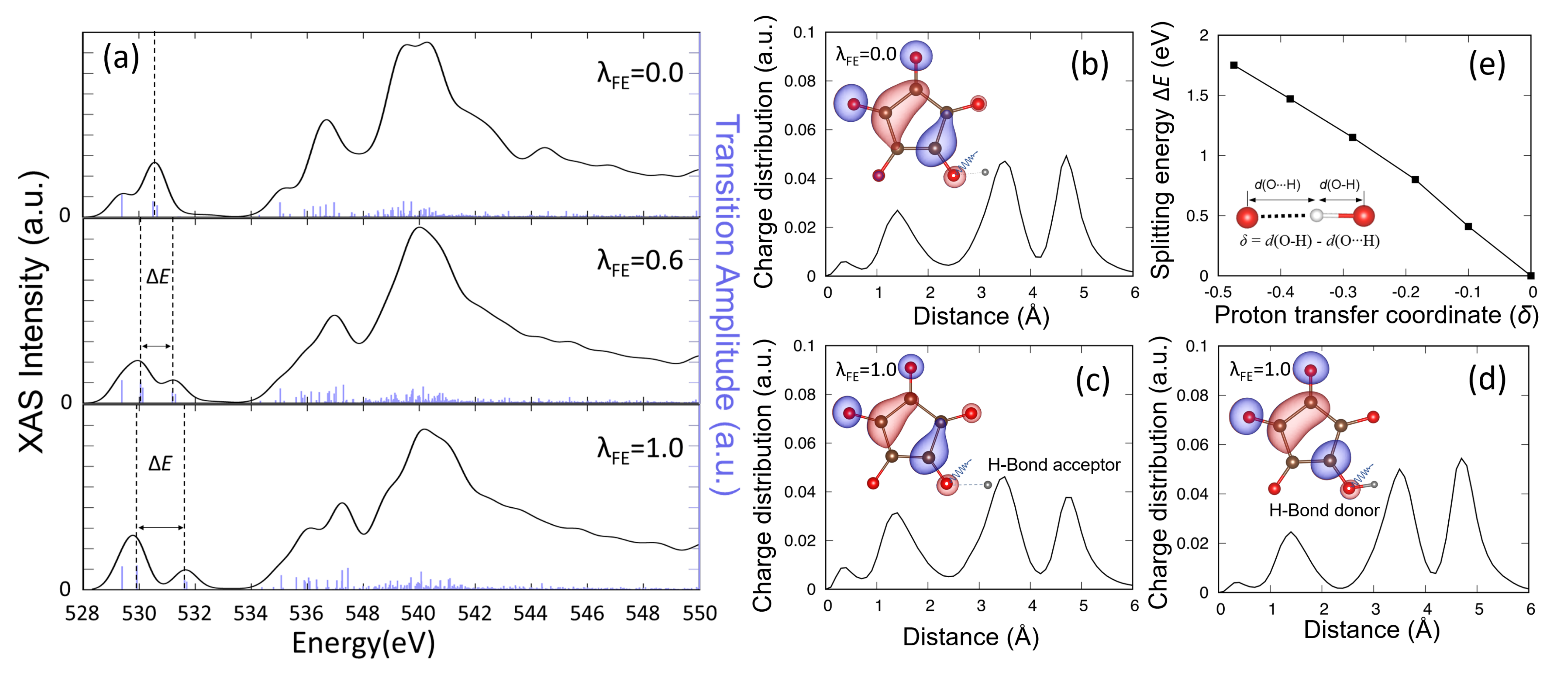}
	\caption{\label{fig:XAS_decomposition}
		(a) The calculated XAS spectra (black) and XAS transition amplitude plots (purple) for CA crystal with different magnitudes of FE distortions of $\lambda_{\rm FE}=0, 0.6, \rm{and}$ 1. The dashed lines denote the peak positions when excited oxygen atoms are in different H-bonding environment. Note that the peak at around $529.4$ eV belongs to the isolated Oxygen atom, whose peak position keeps the same with FE develop. (b), (c) and (d) Charge distribution of the excited oxygen atoms ({\it hinge} direction) where the excited oxygens are oxygen atom in the paraelectric phase, H-bond acceptor pair oxygen atom in the FE phase, and H-bond donor pair oxygen atom in the FE phase, respectively. The zero point are located at the excited oxygen atom. The inserted figures in (b), (c) and (d) are selected excited molecular orbitals, where the white dot indicts the excited oxygen atoms. (e) The relation between splitting energy $\rm {\Delta}$E and proton transfer coordinate $\rm {\delta}$. The inserted figure is the schematic of the definition of proton transfer coordinate.}
\end{figure*}

 Accurate prediction of the XAS spectrum allows us to further explore the spectral
signature of FE in CA. To this end, we rely on theory to study
the spectral changes when the FE ordering is artificially varied in the crystal.
Specifically, we start from the centrosymmetric phase with zero FE distortion
$\lambda_{\rm FE}=0$, and then gradually increase the polarity via $\lambda_{\rm FE}$ in the structure.
In the above,  the $\lambda_{\rm FE}=1$ is normalized to the amplitude of polar mode $\Gamma_2^{-}$,
which is extracted from fully relaxed structure of CA crystal
at ground state of $T$ = 0 K facilitated by group theory analysis \cite{Kroumova2003}.
The resulting XAS spectra for CA crystal with different magnitudes of FE distortions of
$\lambda_{\rm FE}=0, 0.6, \rm{and}$ 1 are shown in Fig.~\ref{fig:XAS_decomposition}(a).
It can be seen that the main-edge of XAS is rather insensitive to the presence of FE ordering.
The general line shape of main-edge remains more or less unaffected when the FE distortion is varied in the crystal.
In sharp contrast, the pre-edge of XAS undergoes qualitative changes upon the development of FE distortion.
In the hypothetical paraelectric structure, only one broad feature is observed in the pre-edge.
Once the FE distortion is added into the hypothetical paraelectric crystal structure, a second peak emerges at slightly higher energy.
With the increasing magnitude of polar distortions, the satellite peak continues moving toward higher energies until
it evolves into an experimentally observed feature at 533 eV in Fig.~\ref{fig:XAS_overall} when $\lambda_{\rm FE}=1$.
Clearly, the appearance of the satellite peak in pre-edge signifies the development of FE
in the material. In XAS spectroscopy, the oxygen $1s$ electron at core level is excited probing the
$p$ character in the unoccupied bands. Therefore, such a feature should be closely associated
with the orbital assignment of the excitons.

\par We assign the pre-edge to a bound exciton which is strongly localized on the
excited molecule in the crystal \cite{Hahner2006c}. The exciton in pre-edge is $\pi^{\ast}$ state which is
antibonding in nature as shown in the insert of Fig.~\ref{fig:XAS_overall}.
The $\pi^{\ast}$ characteristic can be traced back to the low-lying states of molecular excitation
with similar orbital symmetry in the gas phase of CA. In the $\pi^{\ast}$ orbital on the CA,
$p$ character can be identified on the excited oxygen atom which gives rise to a strong transition
amplitude that is responsible for the pre-edge feature.
Moreover, in condensed phase, the antibonding orbital will be influenced
by the local H-bonding environment in a nontrivial way.

\par For each C$_5$H$_2$O$_5$ molecule in the crystal, four of the carbon bonded oxygen atoms will be
also H-bonded to neighboring molecules through the intermolecular protons.
In the centrosymmetric phase with $\lambda_{\rm FE}=0$, the proton is right in the middle of the two molecules.
As a result, the symmetric H-bonding environments are formed on each of the two bonded molecules.
For the molecules bonded along the aforementioned {\it terrace} and {\it hinge} directions,
their H-bonding strengths are rather similar because of the similar intermolecular distances
around $2.41$ \AA. As required by the inversion symmetry, exactly the same  $\pi^{\ast}$
excitonic state, therefore the same transition matrix elements will be generated
for any one of the H-bonded pair oxygen atoms being excited
in the pre-edge as shown in Fig.~\ref{fig:XAS_decomposition} (b).
As a result, a sharp peak at 530.6 eV is formed in pre-edge of XAS, which originates from
the almost degenerate XAS transition amplitudes for all the excited oxygen atoms
in symmetric H-bonding environment.
Besides the H-bonded oxygen atoms, there is also one non-bonded oxygen atom in each
C$_5$H$_2$O$_5$ molecule. Without accepting a proton, the bare lone electron pair generate
more attractive core-hole potential when it is excited in XAS.
Indeed, compared to that on the bonded oxygens, the $\pi^{\ast}$ excitonic state localized on the
non-bonded oxygen has a slightly lower transition energy at 529.4 eV in Fig.~\ref{fig:XAS_decomposition} (a).
Nevertheless, the feature of pre-edge is mainly determined by H-bonded oxygen atoms
in the hypothetical paraelectric phase. The pre-edge is skewed to the left due to the exciton
on the non-bonded oxygen atoms.

\par By including FE ordering into the crystal structure with $\lambda_{\rm FE}=$ 0.6 and 1.0,
the inversion symmetry is broken. As described above, FE distortion is accompanied by large proton transfer.
For the two bonded molecules, proton transfer drives a proton to leave away
from one molecule but get close to the other.
Obviously, the two molecules are not in equivalent H-bonding environments any longer;
instead they become H-bond acceptor and H-bond donor respectively
as shown in Fig.~\ref{fig:XAS_decomposition} (c) and (d).
By carrying out positive electron charge, the proton displacement in turn affects the
energy and localization of bound $\pi^{\ast}$ exciton in the pre-edge.
As far as the H-bond acceptor molecule is concerned, the increased distance between the oxygen and
proton leaves the excited oxygen with a stronger core-hole potential which is less affected by the
Coulombic attraction of the proton. Therefore, the $\pi^{\ast}$ exciton is more localized on the excited oxygen
in Fig.~\ref{fig:XAS_decomposition} (c), which enhances the absorption intensity
with a slightly lower transition energy.
On the contrary, the opposite effect occurs on the H-bond donor molecule.
The localization of bound $\pi^{\ast}$ exciton, on the excited oxygen, is diminished
in Fig.~\ref{fig:XAS_decomposition} (d) due to the increased Coulombic interaction from the closer proton.
Consistently, the absorption intensity is decreased and shifted to a higher transition energy
as shown in Fig.~\ref{fig:XAS_decomposition} (a).
As a result, the development of FE splits the pre-edge resulting in a satellite peak at higher
excitation energy. Furthermore, the splitting energy ${\Delta}$E is roughly linearly proportional to the magnitude of
proton transfer as measure by the proton transfer coordinate $\delta$ in Fig.~\ref{fig:XAS_decomposition} (e).
The above is consistent with the first order approximation of electrostatic energy changes
which is perturbed by the proton transfer.

\par Compared to the bound exciton in the pre-edge features, the main-edge should be assigned to
the exciton resonance states as shown in Fig.~\ref{fig:XAS_overall} \cite{Hahner2006c}. In particular,
two main orbital characteristics are identified here. The peak region around $540$ eV
is attributed to the exciton resonance states with antibonding $\sigma^{\ast}$ character. While the kink feature
is due to the transitions from exciton resonance states with $\pi^{\ast}$ orbitals on the excited molecule,
which are hybridized with $\sigma^{\ast}$ on the surrounding molecules \cite{Hahner2006c}.
Because of their unbound nature, these exciton resonance states are strongly delocalized over the entire
crystal. Therefore, the observed feature in the main-edge is not from the large transition matrix element but
due to the electronic density states in excited states.
Not surprisingly, these delocalized orbitals are not sensitive to the variation
of FE ordering, which applies a local change on the H-bonding environment
in the vicinity of excited oxygen atoms. Indeed, the above is consistent with
the fact that the general main-edge feature is barely changed in Fig.~\ref{fig:XAS_decomposition} (d)
under the artificially adjusted magnitudes of polar distortion $\lambda_{\rm FE}$ in the molecular crystal.

\section{CONCLUSION}

\par In conclusion, we demonstrate that X-ray absorption spectroscopy can be employed as
a sensitive local probe for broken spatial inversion symmetry in CA, which is a prototypical
H-bonded organic molecular crystal. The emergence of a satellite feature in the pre-edge
with a relatively higher transition energy can be denoted as an unambiguous signature of FE.
It arises from the bound $\pi^{\ast}$ exciton whose locality and energy are affected by the two
inequivalent H-bonding configurations distinctly as a result of inversion symmetry breaking.
A similar scenario facilitated by proton transfer lies at the heart of FE in all the H-bonded molecular
ferroelectrics, therefore our approach should be generally applicable to the new series of
organic ferroelectric materials.

\begin{acknowledgments}
\par The authors thank Dr. Pratikkumar Dhuvad and Dr. Zhaoru Sun for helpful discussions. This work was primarily supported by National Science Foundation through Awards No. DMR-1552287 (F. T. and X. W.). X. J. and X.X. acknowledge the support from the U.S. Department of Energy (DOE), Office of Science, Basic Energy Sciences (BES), under Award No. DE-SC0019173 (sample fabrication). G. H. and P. A. D. acknowledge the support from the National Science Foundation through Awards No. Chem-156592 (X-ray spectroscopy). The computational work used resources of the National Energy Research Scientific Computing Center (NERSC), a U.S. Department of Energy Office of Science User Facility operated under Contract No. DE-AC02-05CH11231. This research includes calculations carried out on Temple University's HPC resources and thus was supported in part by the National Science Foundation through major research instrumentation grant number 1625061 and by the US Army Research Laboratory under contract number W911NF-16-2-0189." Use of the Advanced Light Source was supported by the U.S. Department of Energy, Office of Science, Office of Basic Energy Sciences under Contract No. DE-AC02-05CH11231. The research was performed in part in the Nebraska Nanoscale Facility: National Nanotechnology Coordinated Infrastructure and the Nebraska Center for Materials and Nanoscience, which are supported by the National Science Foundation under Grant No. ECCS-1542182, and the Nebraska Research Initiative. Research is carried out in part at the Center for Functional Nanomaterials, which is a U.S. DOE Office of Science Facility, and the Scientific Data and Computing Center, a component of the Computational Science Initiative, at Brookhaven National Laboratory under Contract No. DE-SC0012704.
\end{acknowledgments}

\bibliography{croconic_acid_xas}

\end{document}